\documentclass[12pt]{article}

\usepackage{amssymb,amsmath}

\usepackage{graphicx}
\DeclareGraphicsExtensions{.pdf,.png,.jpg}

 \catcode`\@=11 \@addtoreset{equation}{section}\catcode`\@=12

\begin{document}

 \begin{center}

 \large \bf Stable exponential cosmological
solutions with three factor spaces of dimensions $m=3$, $k_1=k$  and $k_2 = k$  in the Einstein-Gauss-Bonnet model  with a $\Lambda$-term
  \end{center}

 \vspace{0.3truecm}

 \begin{center}

  K. K. Ernazarov

\vspace{0.3truecm}

  \it  Institute of Gravitation and Cosmology,
   Peoples' Friendship University of Russia (RUDN University),
   6 Miklukho-Maklaya St.,  Moscow 117198, Russian Federation


\end{center}

\begin{abstract}
We consider a $(4 + 2k)$ - dimensional  Einstein-Gauss-Bonnet model with the cosmological $\Lambda$-term.
 Exact stable solutions with three constant  Hubble-like parameters in this model are obtained. In this case, the multidimensional cosmological model deals with  three factor spaces: the external   3-dimensional ``our'' world and internal subspaces with dimensions $k_1 = k$ and $k_2=k$. 
\end{abstract}

\section{Introduction}

\indent

In this paper we consider a $D$-dimensional gravitational model
with Gauss-Bonnet term and cosmological term $\Lambda$, which
extends the model with cosmological $\Lambda$  term. 

In modern theoretical cosmology there are two important problems. The first problem is the “dark side” of our Universe. Recent astronomical observations show that our Universe is spatially flat with $\sim 23 \%$  of its critical energy in non-relativistic cold dark matter and $ \sim 73 \% $ having a large negative pressure (dark energy). The existence of dark energy calls the acceleration of the expansion of our Universe, which continues to the present. The second important problem is the possibility of the existence of additional dimensions of the Universe, which follows from string theories. The idea of multidimensionality of our Universe, which follows from the strategy of  unified theory of fundamental interactions, is one of the most interesting ideas of theoretical physics. The idea has its origin from the innovative works of T. Kaluza and O. Klein. In 1919,  once after the creation of the general relativity, Theodore Kaluza proposed to geometrize the electromagnetic field in the spirit of Einstein theory by increasing the number of spatial dimensions by one. In O. Klein's ideas, the 5-dimensions was applied to generalize the relativistic wave equation for a massive particle to the case of a 5-dimensional theory. And now the most self-consistent modern theories, such as superstrings, supergravity, and M-theory, are built on the basis of the theory of multidimensional space-time. Various fruitful ideas of multidimensional gravity are widely used in numerous modern works. So, there is a promising assumption to explain dark energy and the accelerated expansion of our Universe with the help of extra dimensions. It is well known that the dynamical behavior of internal spaces usually leads to variations in the effective gravitational constant (\cite{Zhuk} - \cite{Alimi}) and references therein. Such variations have strong experimental estimates \cite{Uzan}. So, one of the main problems of more voluminous models is the stable compactification of interior spaces. The scale factors of internal spaces play the role of scalar fields moving in our four-dimensional space-time. In this paper, we consider a nonlinear gravitational multidimensional cosmological model with the action (\ref{2.3A}) given below. Each of the spaces has its own scale factor.

\section{The setup}

\indent 

Here  the metric 

\begin{equation}
 g= w du \otimes du + \sum_{i=1}^{n} e^{2\beta^i(t)}\epsilon_i  dy^i \otimes dy^i.
 \label{2.1A}
\end{equation}
is defined on the manifold
\begin{equation}
   M = R  \times   M_1 \times \ldots \times M_n 
   \label{2.2A},
\end{equation}
where $ w=\pm1$, $\epsilon_i=\pm1$, $ i=1, ... , n$. The manifold M is defined as product of one-dimensional manifolds $M_1, ... , M_n$.
The functions $ \gamma(u) $ and $\beta^i(u)$  are smooth  in the open real set $R_*=(u_-, u_+) $. The metric  (\ref{2.1A})  is cosmological for $w=-1$, $\epsilon_1=\epsilon_2= ...=\epsilon_n=1$  and for physical applications the manifolds $M_1$, $M_2$ and $M_3$  are equal to $\mathbb{R}$  and the other manifolds are considered as compact sets.

In our model, the action is expressed as
\begin{equation}
  S =  \int_{M} d^{D}z \sqrt{|g|} \{ \alpha_1 (R[g] - 2 \Lambda) +
              \alpha_2 {\cal L}_2[g] \},
 \label{2.3A}
\end{equation}
where $g=g_{MN}dz^M \otimes dz^N$ is the metric defined on the manifold M, $dimM=D$, $|g|=|det(g_{MN}| $ and

\begin{equation}
 {\cal L}_2 = R_{MNPQ} R^{MNPQ} - 4 R_{MN} R^{MN} +R^2
 \label{2.4A}
\end{equation}
is the standard Gauss-Bonnet term and $\alpha_1$, $\alpha_2$ are nonzero constants. Further we denote  $\alpha= \frac{\alpha_2}{\alpha_1}$. 

Today in cosmology, active research is being conducted based on experimental data. The main goal of research in cosmology is the correct description of the evolution of the Universe since the (hypothetical) ``big bang''. Certain results have been obtained in the field of the study of the evolution of the (so-called) ``early universe''. Various experimental results have confirmed that inflation accurately describes these early stages of evolution. Currently there are sufficiently convincing arguments that explain the accelerated expansion of the universe. The mathematical description of cosmology is provided by the Einstein equations. The introduction of branes into cosmology has given another new approach to our understanding of the Universe and its evolution. In cosmology on branes, it is proposed that our observable universe is a three-dimensional surface embedded in a space of higher dimension.In cosmology, branes need to consider gravity in higher dimensions. Based on Gauss-Bonnet gravity, another consistent multidimensional theory of gravity
with a more general action is obtained, and this action is expressed using string theory.

In this case, we need solutions of multi-dimensional Gauss-Bonnet gravity in the bulk space-time. The Gauss-Bonnet term added to the Einstein-Hilbert term gives the most common action in a multidimensional model with second-order field equations, as shown by Lovelock \cite{Lovelock_1}. 
In multidimensional theories governed only by metric GB term is present  in the bulk  action which is presented by formula (\ref{2.3A}) below. In the classical Gauss-Bonnet theory, $\alpha$ can have any sign. It was shown in \cite{Davis_2} that, for the Gauss – Bonnet worlds, a negative $\alpha$ value leads to the appearance of antigravity or tachyon modes on the brane. 
The cosmology of Gauss-Bonnet gravity has been studied in detail in a number of papers and at present such exact analytical solutions as cosmological  (\cite{Pavl_Mak_Topor_1} - \cite{IvKob-19-1}), centrally symmetric (generalization of Schwarzschild metric based on the Einstein-Gauss-Bonnet gravity) (\cite{R.Kon_2A} - \cite{Torii}) and wormhole (\cite{Kanti} - \cite{Cu_Kon_Zhid}) solutions have been obtained. The  Gauss-Bonnet gravity has been improved in view of physical applications. In ref. \cite{Benetti_18_1}, observational constraints on the Gauss-Bonnet  gravity are discussed in view of Planck satellite data. Lagrange multipliers for the Gauss-Bonnet  gravity are considered in ref. \cite{Sal_Cap_13_1}. Cosmological inflation in the theory of Gauss-Bonnet gravity $f(R, G)$ is discussed in ref. \cite{Laurentis_15_1}.

Here we are dealing with the cosmological  solutions with diagonal metrics (of Bianchi-I-like type) governed by $n$ scale factors depending upon one variable, where $n>3$. Morever, we restrict ourselves by the solutions with exponental dependence of scale factors (with respect to synchronous variable $t$).
\begin{equation}
  a_i(t)  \sim exp(h^i t) 
   \label{2.5A},
\end{equation}
$i=1, ... ,n; $ $D = n+1$.

Latest astronomical observations allege that our Universe is expanding with acceleration. The 3-
dimensional exponential expansion of the Universe is described by following Hubble-like parameters:

\begin{equation}
  h^1=h^2=h^3=H>0
   \label{2.6A}.
\end{equation}

The integrand in (2.3), by substituting the metric (2.1) ,  is expressed as a follow:
\begin{equation}
 \sqrt{|g|}\Bigl\{\alpha_1(R[g]-2\Lambda) +\alpha_2 {\cal L}_2[g]\Bigl\}=L+\frac{df}{du}
   \label{2.7A},
\end{equation}
where
\begin{equation}
 L=\alpha_1 L_1+\alpha_2 L_2
   \label{2.8A}.
\end{equation}

Here terms $L_1$ and $L_2$  read  as follows \cite{IvKob-19-1}:
\begin{equation}
  L_1=(-w)e^{-\gamma+\gamma_0}G_{ij}\dot\beta^i\dot\beta^j - 2 \Lambda e^{\gamma+\gamma_0}
   \label{2.9A},
\end{equation}
\begin{equation}
 L_2=-\frac{1}{3}e^{-3\gamma+\gamma_0}G_{ijkl}\dot\beta^i\dot\beta^j\dot\beta^k\dot\beta^l
   \label{2.10A},
\end{equation}
where
\begin{equation}
 \gamma_0= \sum_{i=1}^n \beta^i
   \label{2.11A}.
\end{equation}

Here a 2-linear symmetric form is used in the “mini-supermetric” - 2 - metric of pseudo-Euclidean signature:
\begin{equation}
<\upsilon_1, \upsilon_2> = G_{ij}\upsilon_1^i\upsilon_2^j
   \label{2.12A},
\end{equation}
where
\begin{equation}
 G_{ij}= \delta_{ij}-1 
   \label{2.13A},
\end{equation}
and  a 4-linear symmetric form -  Finslerian 4-metric:
\begin{equation}
<\upsilon_1, \upsilon_2, \upsilon_3, \upsilon_4> = G_{ijkl}\upsilon_1^i\upsilon_2^j\upsilon_3^k\upsilon_4^l
   \label{2.14A}
\end{equation}
with components 
\begin{equation}
 G_{ijkl}= (\delta_{ij}-1) (\delta_{ik}-1) (\delta_{il}-1) (\delta_{jk}-1) (\delta_{jl}-1) (\delta_{kl}-1) 
   \label{2.15A}.
\end{equation}

 The function $f(u)$ in  (\ref{2.7A})  is irrelevant for our consideration (see \cite{Iv-09}, \cite{Iv-10}) and  we denote $ \dot A=\frac{dA}{du} $.

With help of  following identities (\cite{Iv-09}, \cite{Iv-10}) we can derive (\ref{2.8A}) -  (\ref{2.10A}):
\begin{equation}
 G_{ij}\upsilon^i\upsilon^j= \sum_{i=1}^n (\upsilon^i)^2 - \Biggl(\sum_{i=1}^n \upsilon^i\Biggl)^2 
   \label{2.16A},
\end{equation}

\begin{eqnarray}
 G_{ijkl}\upsilon^i\upsilon^j\upsilon^k\upsilon^l= \Biggl( \sum_{i=1}^n \upsilon^i \Biggl)^4 - 6\Biggl(\sum_{i=1}^n \upsilon^i\Biggl)^2\sum_{j=1}^n (\upsilon^j)^2   \nonumber \\
+  3\Biggl(\sum_{i=1}^n (\upsilon^i)^2\Biggl)^2 + 8\Biggl(\sum_{i=1}^n \upsilon^i\Biggl) \sum_{j=1}^n (\upsilon^j)^3 - 6\sum_{i=1}^n (\upsilon^i)^4 
   \label{2.17A}.
\end{eqnarray}

The following form of the equation of motion is got  from the action  (\ref{2.3A}):

\begin{equation}
 \epsilon_{MN} = \alpha_1\epsilon_{MN}^{(1)} + \alpha_2\epsilon_{MN}^{(2)} = 0 
   \label{3.1A},
\end{equation}

where

\begin{equation}
 \epsilon_{MN}^{(1)} = R_{MN} - \frac{1}{2}Rg_{MN} + \Lambda 
   \label{3.2A},
\end{equation},

\begin{equation}
 \epsilon_{MN}^{(2)} = 2 \Biggl(R_{MPQS}R_N^{PQS} - 2R_{MP}R_N^P - 2R_{MPNQ}R^{PQ} + RR_{MN} \Biggl) - \frac{1}{2}{\cal L}_2 g_{MN}
   \label{3.3A}.
\end{equation}

Now we put $w= -1$,  and the follow set of  polynomial equations is taken from  the equations of motion for the action (\ref{2.3A})
\begin{eqnarray}
  E = G_{ij} v^i v^j + 2 \Lambda
  - \alpha   G_{ijkl} v^i v^j v^k v^l = 0,  \label{3.4A} \\
   Y_i =  \left[ 2   G_{ij} v^j
    - \frac{4}{3} \alpha  G_{ijkl}  v^j v^k v^l \right] \sum_{i=1}^n v^i 
    - \frac{2}{3}   G_{ij} v^i v^j  +  \frac{8}{3} \Lambda = 0,
   \label{3.5A}
\end{eqnarray}

$i = 1,\ldots, n$, where  $\alpha = \alpha_2/\alpha_1$. 
A set of forth-order polynomial  equations is got for $n > 3$. 

In the set of equations (\ref{3.4A}) and (\ref{3.5A}) can be found  an isotropic solution $v^1 = \cdots = v^n = H$  for $\Lambda =0$ and $n > 3$ only if $\alpha  < 0$ \cite{Iv-09,Iv-10}. This solution was spread in \cite{ChPavTop} to the case $\Lambda \neq 0$.

In the articles \cite{Iv-09,Iv-10} was shown  that there are no more than
three different  numbers among  $v^1,\dots ,v^n$ when $\Lambda =0$. And it was proven that it  is valid also
for  $\Lambda \neq 0$ if $\sum_{i = 1}^{n} v^i \neq 0$. 

\section{Exponential solutions with the  three Hubble-like parameters}

\indent

We start this section with more general task (which will be a subject of another paper), i.e the task to find a class  of solutions to the set of equations (\ref{3.4A}), (\ref{3.5A}) of the following form:
\begin{equation}
  \label{3.1}
   v =(\underbrace{H,H,H}_{``our'' \ space},\underbrace{\overbrace{h_1, \ldots, h_1}^{k_1}, \overbrace{h_2, \ldots, h_2}^{k_2}}_{internal \ space}),
\end{equation}
where $H$ is the Hubble-like parameter corresponding to an $3$-dimensional  factor  space. The Hubble-like parameters of  $k_1$-dimensional and $k_2$-dimensional factor spaces  with $k_1 > 1$ and  $k_2 > 1$ are  accordingly  $h_1$  and $h_2$. Note that in our considered model $h_2 \neq h_1$.  The Hubble-like parameter  $H$ describes expanding of  "our" $3d$ space while the next parameters $h_1$  and $h_2$ describe expanding or contraction of $( k_1 + k_2)$-dimensional internal space.

We assume  
\begin{equation}
  \label{3.2a}
   H > 0 
\end{equation}
for a description of an accelerated expansion of a
$3$-dimensional subspace (which may describe our Universe).

The (so-called) "our"  $3$-dimensional factor space is expanding with the Hubble parameter $H >0$, while the $k_i$-dimensional internal  factor space  is contracting with the Hubble-like  parameter $h_i < 0$, where $i$ is 
either $1$ or $2$  as shown in the ansatz (\ref{3.1}).

Then we consider the ansatz (\ref{3.1}) with Hubble-like parameters $H$, $h_1$ and $h_2$ which imposed the following restrictions:
\begin{equation}
   S_1 = 3 H +  k_1h_1 + k_2 h_2 \neq 0, \quad  H \neq h_1, \quad  H \neq h_2, \quad  h_1 \neq h_2, \quad  k \neq 1.
   \label{3.3}
   \end{equation}

 It was shown in ref. \cite{ErIv-17-2} that  the set of $(n + 1)$ polynomial equations  
 (\ref{3.4A}), (\ref{3.5A}) with  ansatz  
 (\ref{3.1}) and restrictions (\ref{3.3}) imposed  is reduced to a set  of fourth,  second and first order polynomial equations,  accordingly:
    
    \begin{eqnarray}
          E =0,   \label{3.4E} \\
          Q =  - \frac{1}{2 \alpha}, \label{3.4Q} \\
          L = H + h_1 + h_2 - S_1 = 0.  \label{3.4L}
     \end{eqnarray}
   where  $E$ is defined in (\ref{3.4A})  with ($v^1$,$v^2$,$v^3$) = (H, $h_1$, $h_2$) and 
   \begin{equation}
        Q = Q_{h_1 h_2} =  S_1^2 - S_2 - 2 S_1 (h_1 + h_2) + 2 (h_1^2 + h_1 h_2 + h_2^2),
                   \label{3.5}
        \end{equation}
$S_1$  is defined in (\ref{3.3})  and $ S_2 = 3 H^2 + k_1 (h_1)^2 + k_2 (h_2)^2$
     
 As it was proved in \cite{ErIv-17-2}  by using results of ref. \cite{Ivas-16-1} (see also \cite{Pavl-15}),  the
 exponential solutions with  $v$ from (\ref{3.1}) and $k_1 > 1$, $k_1 > 2$ are stable if and only if 
 \begin{equation}
           S_1 = 3 H +  k_1h_1 + k_2 h_2 = H + h_1 + h_2 > 0. 
                       \label{3.6}
 \end{equation}
Above we use the relation (\ref{3.4L}).

\subsection{Exact stable solutions in $(3+ 2 k)$-dimensional case}

\indent

In our further research, the solutions to the set of equations of motion are presented in the form:
\begin{equation}
v =(\underbrace{H,H,H}_{''our'' space},\underbrace{\overbrace{h_1, \ldots, h_1}^k, \overbrace{h_2, \ldots, h_2}^{k}}_{internal \ space})
\label{6.1A}
\end{equation}

\noindent 
where $k_1 = k_2 = k > 3$  and $H$ is the Hubble-like parameter that corresponds to the 3-dimensional “our” subspace and the $h_1$, $h_2$ are Hubble-like parameters that correspond to the internal subspaces of dimensions
$k_1$ and $k_2 $, respectively.
The following conditions must be imposed on the solutions:

A) $H>0$. In this model "our"  3-dimensional world is expanding with acceleration. Therefore, the Hubble-like  parameter corresponding to this world should be positive. The remaining dimensions are considered as an internal subspace dimensions.

B) And it can also be noted that our considered model is anisotropic. Therefore, there is an unambiguous expansion in “our” 3-dimensional world and the rest of the internal dimensions either have a contraction, or an expansion in some dimensions and a contraction in the other internal dimensions. From such considerations follows the need to comply with next condition: 

\noindent B.1) ($h_1 < 0$, $h_2 < 0$) - a contraction in the internal subspace; \newline
B.2) ($h_1 < 0$, $h_2 > 0$) - a contraction in the internal $k_1$-dimensions and an  expansion in the internal $ k_2 $-dimensions; \newline
B.3) ($h_1 > 0$, $h_2 < 0$) - an expansion in the internal $k_1$-dimensions and a contraction in the internal $k_2$ -dimensions. 

We note that the solutions with $H>0$, $h_1>0$, $h_2 >0$ do not  appear in our consideration due to 
relation (\ref{6.5A}).

Once our results have complied with the above conditions, from  (\ref{3.4Q}) the following solutions are obtained  in case $(m, k_1, k_2) = (3, k, k)$:

\begin{equation}
h_1 = -\frac{H\pm \sqrt{\frac{k-1}{4\alpha} - kH^2 }}{k-1}
   \label{6.2A}.
 \end{equation}

By substitution of  (\ref{6.2A}) and $\lambda=\Lambda\alpha$, $X=\alpha H^2$  into relation  (\ref{3.4E})  we get the following expression: 
\begin{equation}
AX^2 + BX + C = 0
   \label{6.3A},
 \end{equation}
here

\begin{equation}
A= 4k(k + 1)(k - 3)
   \label{6.3AA},
 \end{equation}

\begin{equation}
B= -2(k-1)(k-3)
   \label{6.3AB}
 \end{equation}

and 
\begin{equation}
C = \frac{k(k-1)^2}{4} - 2\lambda(k-1)^3
   \label{6.3C}.
 \end{equation}

We find the roots of the quadratic equation and get the following results:

\begin{equation}
X = \alpha H^2 = - \frac{B \pm \sqrt{B^2 - 4AC}}{2A}
   \label{6.4A}.
 \end{equation}

Thus, the last expressions show that all the Hubble - like  parameters $H $ and $h_1$  are uniquely determined by $k$ and $\lambda$. The definition of the Hubble-like  parameter $h_2$  in the considered multidimensional model is carried out using equation  (\ref{3.4L}):

\begin{equation}
h_2 = - \Biggl(\frac{2}{k-1}H + h_1\Biggl)
   \label{6.5A}.
 \end{equation}

The relation  
\begin{equation}
S_1 =  \frac{(k- 3 )}{k-1} H > 0.
\label{6.6B}
\end{equation}
provides us with an opportunity to select the stable solutions.
For $k > 3$ we have stable solutions for $H > 0$ and unstable  - 
for $H < 0 $.

\section{Examples}

\subsection{ $k_1 =  k_2 = k = 5$  and $\alpha > 0$}

\indent

Let us consider the case  $ k = 5$. From (\ref{6.4A}) we get 

 \begin{eqnarray}
H = \frac{1}{\sqrt{30\alpha}}\sqrt{1 \pm \sqrt{480\lambda - 74}} \nonumber \\
H = - \frac{1}{\sqrt{30\alpha}}\sqrt{1 \pm \sqrt{480\lambda - 74}}
   \label{7.1A}
 \end{eqnarray}

\noindent and further, as our calculations show, each value of four solution of $H$ corresponds to two values of the solution $h_1$ (see (\ref{6.2A})) and one number of the solutions $h_2$  (see (\ref{6.5A})). Therefore, the eight number of the set of real solutions can be found and as it is shown in our calculations, among of them four solutions are unstable.
 Therefore, the imposition of  the stability condition and the conditions A), B)  reduce the number of the set of stable real solutions to four:

1) 

 \begin{equation}
H = \frac{1}{\sqrt{30\alpha}}\sqrt{1 + \sqrt{480\lambda - 74}}
   \label{7.2A},
 \end{equation}

  \begin{equation}
h_1 =  - \frac{1}{4\sqrt{30\alpha}}\Biggl(\sqrt{1 + \sqrt{480\lambda - 74}} -  \sqrt{25 - 5\sqrt{480\lambda - 74}}\Biggl)
   \label{7.3A},
 \end{equation}

 \begin{equation}
 h_2 = - \frac{1}{4\sqrt{30\alpha}}\Biggl(\sqrt{1 + \sqrt{480\lambda - 74}} + \sqrt{25 - 5\sqrt{480\lambda - 74}}\Biggl)
   \label{7.4A},
 \end{equation}

 \begin{equation}
S_1 = \frac{1}{2\sqrt{30\alpha}}\sqrt{1 + \sqrt{480\lambda - 74}} > 0
   \label{7.4AB}.
 \end{equation}

The interval of $\lambda$, which occured the solutions $H > 0$, $h_1 < 0$ and $h_2 < 0 $  is:

\begin{displaymath} 
\frac{30}{160} < \lambda < \frac{33}{160}
\end{displaymath} 

and the solutions $H > 0$, $h_1 > 0$ and $h_2 < 0 $ are existed in the interval of $\lambda$: 
\begin{displaymath} 
\frac{37}{240} < \lambda < \frac{30}{160}.
\end{displaymath} 

2)

 \begin{equation}
H = \frac{1}{\sqrt{30\alpha}}\sqrt{1 + \sqrt{480\lambda - 74}}
   \label{7.2AC},
 \end{equation}

  \begin{equation}
h_1 =  - \frac{1}{4\sqrt{30\alpha}}\Biggl(\sqrt{1 + \sqrt{480\lambda - 74}} + \sqrt{25 - 5\sqrt{480\lambda - 74}}\Biggl)
   \label{7.3AC},
 \end{equation}

 \begin{equation}
 h_2 = - \frac{1}{4\sqrt{30\alpha}}\Biggl(\sqrt{1 + \sqrt{480\lambda - 74}} - \sqrt{25 - 5\sqrt{480\lambda - 74}}\Biggl)
   \label{7.4AC},
 \end{equation}

 \begin{equation}
S_1 = \frac{1}{2\sqrt{30\alpha}}\sqrt{1 + \sqrt{480\lambda - 74}} > 0
   \label{7.4ABF}.
 \end{equation}

The interval of $\lambda$, which occured the solutions $H > 0$, $h_1 < 0$ and $h_2 < 0 $  is:

\begin{displaymath} 
\frac{30}{160} < \lambda < \frac{33}{160}
\end{displaymath} 

and the solutions $H > 0$, $h_1 < 0$ and $h_2 > 0 $ are existed in the interval of $\lambda$: 
\begin{displaymath} 
\frac{37}{240} < \lambda < \frac{30}{160}.
\end{displaymath}

3)

 \begin{equation}
H = \frac{1}{\sqrt{30\alpha}}\sqrt{1 -  \sqrt{480\lambda - 74}}
   \label{7.5A}.
 \end{equation}

 \begin{equation}
h_1 =  - \frac{1}{4\sqrt{30\alpha}}\Biggl(\sqrt{1 - \sqrt{480\lambda - 74}} - \sqrt{25 + 5\sqrt{480\lambda - 74}}\Biggl)
   \label{7.6A},
 \end{equation}

 \begin{equation}
h_2 = - \frac{1}{4\sqrt{30\alpha}}\Biggl(\sqrt{1 - \sqrt{480\lambda - 74}}  + \sqrt{25 + 5\sqrt{480\lambda - 74}}\Biggl)
   \label{7.7A},
 \end{equation}

 \begin{equation}
S_1 = \frac{1}{2\sqrt{30\alpha}}\sqrt{1 - \sqrt{480\lambda - 74}} > 0
   \label{7.7AB}.
 \end{equation}

The interval of $\lambda$, which existed the solutions  $H > 0$, $h_1 > 0$ and $h_2 < 0 $ is:

\begin{displaymath} 
\frac{37}{240} < \lambda < \frac{5}{32}
\end{displaymath}

4)

 \begin{equation}
H = \frac{1}{\sqrt{30\alpha}}\sqrt{1 -  \sqrt{480\lambda - 74}}
   \label{7.5AB}.
 \end{equation}

 \begin{equation}
h_1 =  - \frac{1}{4\sqrt{30\alpha}}\Biggl(\sqrt{1 - \sqrt{480\lambda - 74}} + \sqrt{25 + 5\sqrt{480\lambda - 74}}\Biggl)
   \label{7.6AB},
 \end{equation}

 \begin{equation}
h_2 = - \frac{1}{4\sqrt{30\alpha}}\Biggl(\sqrt{1 - \sqrt{480\lambda - 74}} - \sqrt{25 + 5\sqrt{480\lambda - 74}}\Biggl)
   \label{7.7AB},
 \end{equation}

 \begin{equation}
S_1 = \frac{1}{2\sqrt{30\alpha}}\sqrt{1 - \sqrt{480\lambda - 74}} > 0
   \label{7.7ABF}.
 \end{equation}

The interval of $\lambda$, which existed the solutions $H > 0$, $h_1 < 0$ and $h_2 > 0 $ is:

\begin{displaymath} 
\frac{37}{240} < \lambda < \frac{5}{32}
\end{displaymath}

\subsection{$k_1 =  k_2 = k = 6$ and $\alpha > 0$}

\indent

In this case similar calculations are done as in the previous. So, in the set of dimensions $ (m, k_1, k_2 ) = ( 3, 6, 6 )$, solving the set of polynomial equations  (\ref{3.4E}) - (\ref{3.4L}), one can obtain similar formulas and expressions as in the set of dimensions $ (m, k_1, k_2 ) = ( 3, 5, 5 )$. In this set of dimensions from (\ref{7.1A}) we get four real solutions for $H$ and for each value of four solution of $H$ corresponds to two values of the solution $h_1$ (see (\ref{6.2A})) and four numbers of the solutions $h_2$ (see (\ref{6.5A})). Finally, eight number of the set of real solutions are founded. As our calculations show, four of them are unstable.
 Therefore, the imposition the stability condition and the conditions A), B) reduce the number of the set of stable real solutions to four:

1) 

 \begin{equation}
H = \frac{1}{4\sqrt{21\alpha}}\sqrt{10(1 - \sqrt{560\lambda - 83})}
   \label{7.8A},
 \end{equation}

  \begin{equation}
h_1 =  - \frac{1}{12\sqrt{35\alpha}}\Biggl(\sqrt{6(1 - \sqrt{560\lambda - 83})} + 6\sqrt{6 + \sqrt{560\lambda - 83}}\Biggl)
   \label{7.9A},
 \end{equation}

 \begin{equation}
h_2 =  - \frac{1}{12\sqrt{35\alpha}}\Biggl(\sqrt{6(1 - \sqrt{560\lambda - 83})} -  6\sqrt{6 + \sqrt{560\lambda - 83}}\Biggl)
   \label{7.10A}.
 \end{equation}

 \begin{equation}
S_1 = \frac{1}{4\sqrt{35\alpha}}\sqrt{6(1 - \sqrt{560\lambda - 83})} > 0
   \label{7.10AB}.
 \end{equation}

The interval of $\lambda$, which occured the solutions $H > 0$, $h_1 < 0$ and $h_2 > 0 $ is:

\begin{displaymath} 
\frac{83}{560} < \lambda < \frac{84}{560}.
\end{displaymath}

2)

 \begin{equation}
H = \frac{1}{4\sqrt{21\alpha}}\sqrt{10(1 - \sqrt{560\lambda - 83})}
   \label{7.11A},
 \end{equation}

  \begin{equation}
h_1 =  - \frac{1}{12\sqrt{35\alpha}}\Biggl(\sqrt{6(1 - \sqrt{560\lambda - 83})} - 6\sqrt{6 + \sqrt{560\lambda - 83}}\Biggl)
   \label{7.12A},
 \end{equation}

 \begin{equation}
h_2 =  - \frac{1}{12\sqrt{35\alpha}}\Biggl(\sqrt{6(1 - \sqrt{560\lambda - 83})} +  6\sqrt{6 + \sqrt{560\lambda - 83}}\Biggl)
   \label{7.13A}.
 \end{equation}

 \begin{equation}
S_1 = \frac{1}{4\sqrt{35\alpha}}\sqrt{6(1 + \sqrt{560\lambda - 83})} > 0
   \label{7.13AB}.
 \end{equation}

The interval of $\lambda$, which occured the solutions $H > 0$, $h_1 >  0$ and $h_2 < 0 $  is:

\begin{displaymath} 
\frac{83}{560} < \lambda < \frac{84}{560}.
\end{displaymath}

3)

 \begin{equation}
H = \frac{1}{4\sqrt{21\alpha}}\sqrt{10(1 + \sqrt{560\lambda - 83})}
   \label{7.14A},
 \end{equation}

  \begin{equation}
h_1 =  - \frac{1}{12\sqrt{35\alpha}}\Biggl(\sqrt{6(1 + \sqrt{560\lambda - 83})} - 6\sqrt{6 - \sqrt{560\lambda - 83}}\Biggl)
   \label{7.15A},
 \end{equation}

 \begin{equation}
h_2 =  - \frac{1}{12\sqrt{35\alpha}}\Biggl(\sqrt{6(1 + \sqrt{560\lambda - 83})} +  6\sqrt{6 - \sqrt{560\lambda - 83}}\Biggl)
   \label{7.16A},
 \end{equation}

 \begin{equation}
S_1 = \frac{1}{4\sqrt{35\alpha}}\sqrt{6(1 - \sqrt{560\lambda - 83})} > 0
   \label{7.16AB}.
 \end{equation}

The interval of $\lambda$, which existed the solutions $H > 0$, $h_1 < 0$ and $h_2 < 0 $ is:

\begin{displaymath} 
\frac{108}{560} < \lambda < \frac{119}{560}
\end{displaymath} 

and the solutions $H > 0$, $h_1 >  0$ and $h_2 < 0 $ are occured in the interval of $\lambda$: 
\begin{displaymath} 
\frac{83}{560} < \lambda < \frac{108}{560}.
\end{displaymath}

4)

 \begin{equation}
H = \frac{1}{4\sqrt{21\alpha}}\sqrt{10(1 + \sqrt{560\lambda - 83})}
   \label{7.17A},
 \end{equation}

  \begin{equation}
h_1 =  - \frac{1}{12\sqrt{35\alpha}}\Biggl(\sqrt{6(1 + \sqrt{560\lambda - 83})} + 6\sqrt{6 - \sqrt{560\lambda - 83}}\Biggl)
   \label{7.18A},
 \end{equation}

 \begin{equation}
h_2 =  - \frac{1}{12\sqrt{35\alpha}}\Biggl(\sqrt{6(1 + \sqrt{560\lambda - 83})} -  6\sqrt{6 - \sqrt{560\lambda - 83}}\Biggl)
   \label{7.19A},
 \end{equation}

 \begin{equation}
S_1 = \frac{1}{4\sqrt{35\alpha}}\sqrt{6(1 + \sqrt{560\lambda - 83})} > 0
   \label{7.19AB}.
 \end{equation}

The interval of $\lambda$, which existed the solutions $H > 0$, $h_1 < 0$ and $h_2 < 0 $  is:

\begin{displaymath} 
\frac{108}{560} < \lambda < \frac{119}{560}
\end{displaymath} 

and the solutions $H > 0$, $h_1 <  0$ and $h_2 > 0 $ are occured  in the interval of $\lambda$: 
\begin{displaymath} 
\frac{83}{560} < \lambda < \frac{108}{560}.
\end{displaymath}

\section{Simulation by solution with anisotropic fluid}

The  obtained solutions may be simulated by solutions with anisotropic fluid in the $(4+2k)$-dimensional model
with anisotropic fluid instead of Gauss-Bonnet and $\Lambda$ terms.

Thus, here consider the model  with anisotropic fluid. This model is given
by $(4+2k)$-dimensional Einstein-Hilbert equations 

\begin{equation}
R_N^M - \frac{1}{2}\delta_N^M R = \varkappa^2 T_N^M,
  \label{5A.1A}
 \end{equation}
 where $\varkappa^2 >0 $ is $(4+2k)$-dimensional gravitational constant.  

Here the energy-momentum tensor reads
\begin{equation}
(T_N^M) = diag( -\rho, p_{0},p_{0},p_{0}, \underbrace{p_{1}, ... ,p_{1}}_k,\underbrace{p_{2}, ... ,p_{2}}_k ) 
  \label{5A.2A}
 \end{equation}
and the pressures of this anisotropic fluid are proportional to the density, i.e.
\begin{equation}
p_a = w_a \rho 
  \label{5A.3A},
 \end{equation}
where $w_a = const$, $a = 0, 1,2$ and $\rho \neq 0$.

It may be shown by a straightforward verification that with the choice  
   \begin{equation}
    w_a  = 1  - \frac{2 S_1}{B} (h_a - S_1),
     \label{5A.4A}
    \end{equation}
$a = 0, 1,2$, where $h_0 = H$, 
 \begin{eqnarray}
  S_1 = 3H + k (h_1 + h_2)   \label{5A.5A}, \\
  B = 3H^2 + k (h_1^2 + h_2^2) - S_1^2  \label{5A.6A}
 \end{eqnarray}
and 
\begin{equation}
    \rho  =  - \frac{B}{2 \varkappa^2},
     \label{5A.7A}
    \end{equation}
we get the same solution for the  metric with three Hubble-like parameters as was obtained above (in our EGB$\Lambda$ model).
  
It is of interest to generalize this trick to solutions with an anisotropic fluid in the theory of gravity $f(R, G)$ which are considered by many authors. We note, that in the article \cite{Sal_Cap_19_1} it was shown that, in the  $n$-dimensional Friedmann-Robertson-Walker metric, it is rigorously shown that any analytical theory of Gauss-Bonnet  gravity $f(R, G)$  where $R$ is the curvature scalar and $G$ is the Gauss-Bonnet term, can be associated to a perfect-fluid stress-energy tensor. One may think, that in this perspective, dark components of the cosmological Hubble flow can be geometrically interpreted.

\section{Variation of G}


\indent

 Astronomical observations and studies of type Ia supernovae \cite{ Riess, Perlmutter}  indicate that the Universe has been  recently accelerating and decelerating at earlier stages. The Friedmann model of the Universe without a cosmological constant $\Lambda$  and with zero curvature can not explain this evolution of the Universe \cite{Peebles}. The accelerated expansion of the Universe can be explained with the existence of “dark energy” with negative pressure, the simplest possibility is the introduction of a cosmological constant \cite{Sahni}. The modification theory of gravity is an alternative way to the theory of gravity, for example, considering that the effective gravitational constant G changes with time. Recall that the hypothesis of the time variation of the gravitational constant was first expressed in the work of P.A.M. Dirac in the framework of his Large Number hypothesis \cite{Dirac} and later developed in \cite{Brans} in the framework of the alternative theory of gravity. It is worth mentioning that many theoretical approaches, such as multi-dimensional gravity models, string theories or scalar-tensor models of quintessence, contain a built-in mechanism for a possible time variation of the couplings.

The dimensionless parameter of variation of (effective) gravitational constant
(in Jordan frame) \cite{RZ-98,I-96,BIM,Mel} reads

\begin{equation}
  \label{3.var}
{\rm var.} = \frac{\dot{G}}{G H} = - \frac{k_1h_1 + k_2h_2}{H}. 
\end{equation}

So far, experimental data had shown that the variation of the gravitational constant is allowed
at the level of $10^{-13}$ per year and less. Here the following constraint on the magnitude of the dimensionless variation of the
 effective gravitational constant has used:

 \begin{equation}
 \label{5.G1}
  - 0,63 \cdot 10^{-3} < \frac{\dot{G}}{GH} < 1,13 \cdot 10^{-3}.
 \end{equation}
It comes from the most stringent limitation
on $\dot{G}$ obtained by the set of ephemerides \cite{Pitjeva}

\begin{equation}
 \label{5.G2}
          -0,42 \cdot 10^{-13} \ yr^{-1} <  \dot{G}/G  <  0,75  \cdot 10^{-13} \ yr^{-1}
\end{equation}
allowed at 95\% confidence (2$\sigma$) level
and the present value of the Hubble parameter \cite{Ade}
\begin{equation}
 \label{H}
  H_0 =  (67,3 \pm 2,4) \ km/s \ Mpc^{-1} =  (6,878  \pm 0,245) \cdot 10^{-11} \ yr^{-1},
 \end{equation}
  with 95\% confidence level ($2 \sigma$).

In what follows  we denote $\lambda = \Lambda \alpha$, $\alpha >0$. 

For the families of solutions \{(\ref{7.2A}) - (\ref{7.4A})\},  \{(\ref{7.2AC}) - (\ref{7.4AC})\} , \{(\ref{7.5A}) -(\ref{7.7A})\} and \{(\ref{7.5AB}) - (\ref{7.7AB})\} the value of the variation of the effective gravitational constant G  is determined by formula \ref{3.var} and is always equal to the following:

\begin{displaymath} 
{\rm var.} = \frac{5}{2}.
\end{displaymath} 






In case, when $(m, k_1, k_2) = (3, 6, 6)$ for the families of solutions \{(\ref{7.8A}) - (\ref{7.10A})\}, \{(\ref{7.11A}) - (\ref{7.13A})\}, \{(\ref{7.14A}) - (\ref{7.16A})\}  and \{(\ref{7.17A}) - (\ref{7.19A})\} , the value of the variation of the effective gravitational constant G  is is always equal to the following:

\begin{displaymath} 
{\rm var.} = \frac{12}{5}.
\end{displaymath}

\section{Conclusions}

\indent

We have considered the  $(4 + 2k) $-dimensional  Einstein-Gauss-Bonnet (EGB) model
with the $\Lambda$-term.  
By using the  ansatz with diagonal  cosmological  metrics, we have found 
for $D =  4 + 2k $, $\alpha = \alpha_2 / \alpha_1 > 0$ and certain  $\lambda = \alpha\Lambda$   
 a class of solutions with three Hubble-like parameters $H >0$,  
$h_1$, and $h_2$  corresponding to submanifolds of  
dimensions $m=3 $, $k_1 = k$ and $k_2 = k$, respectively. The obtained solutions are exact and stable. 
As we know, stability plays a predominant role in exact solutions of a set of equations. 
Therefore, we assume to study the obtained solutions in our next paper.

 {\bf Acknowledgments}
The publication was prepared with the support of the “RUDN University
Program 5-100”. It was also partially supported by the Russian Foundation
for Basic Research, grant Nr. 19-02-00346.


\end{document}